%% file: main.tex
\documentclass[conference,11pt]{IEEEtran}

\usepackage{array}
\usepackage[utf8]{inputenc}
\usepackage{amsmath,bm}
\usepackage{amsfonts}
\usepackage{amssymb}
\usepackage{amsthm}
\usepackage{graphicx,booktabs,longtable,xcolor}
\usepackage{algorithm,algorithmic}
\usepackage{booktabs}
\usepackage[printonlyused, nolist]{acronym}
\usepackage{trfsigns}
\usepackage{multirow} 
\usepackage{cite}
\usepackage{url}

\usepackage{tikz,pgfplots,pdflscape,pstricks}
\usetikzlibrary{arrows.meta,backgrounds,calc,decorations,patterns,positioning}
\usetikzlibrary{decorations.pathreplacing,angles,quotes,shapes,fit}
\usepgfplotslibrary{groupplots}
\newlength\fheight 
\newlength\fwidth 

\ifCLASSINFOpdf
\else
\fi

\definecolor{LMSred}{rgb}{0.80,0.20,0.20} 
\definecolor{LMSgray}{rgb}{0.60,0.60,0.60}
\definecolor{LMSCyan}{rgb}{0.60,1,1}
\definecolor{LMSMagenta}{rgb}{0.965,0.294,1}
\definecolor{LMSYellow}{rgb}{0.95,0.95,0.0}
\definecolor{LMSOrange}{rgb}{1,0.5,0}
\definecolor{LMSlightblue}{rgb}{0.5,0.8,1}
\definecolor{LMSlightred}{rgb}{1,0.294,0.294}
\definecolor{LMSblue}{rgb}{0.216,0.255,1}
\definecolor{LMSgreen}{rgb}{0.15,0.7,0.15}
\definecolor{LMSlightgreen}{rgb}{0.35,0.90,0.35}
\definecolor{LMSdarkgreen}{rgb}{0.01,0.5,0.01}

\begin{acronym}
\acro{AWGN}{Additive White Gaussian Noise}
\acro{DNN}{Deep Neural Network}
\acro{CNN}{Convolutive Neural Network}
\acro{ICPHM}{International Conference on Prognostics and Health Management}
\acro{SNR}{Signal-to noise-ratio}
\end{acronym}

\pgfplotsset{compat=1.18}
\begin{document}

\title{1-D Residual Convolutional Neural Network coupled with Data Augmentation and Regularization  for the ICPHM 2023 Data Challenge }

\author{Matthias Kreuzer,  Walter Kellermann \\
\textit{Multimedia Communications and Signal Processing, FAU Erlangen-N\"urnberg, Germany} \\
\texttt{\{matthias.kreuzer, walter.kellermann\}@fau.de}
}
\maketitle

\begin{abstract}
In this article, we present our contribution to the \ac{ICPHM} 2023 Data Challenge on Industrial Systems’ Health Monitoring using Vibration Analysis. For the task of classifying sun gear faults in a gearbox, we propose a residual \ac{CNN} that operates on raw three-channel time-domain vibration signals. In conjunction with data augmentation and regularization techniques, the proposed model yields very good results in a multi-class classification scenario with real-world data despite its relatively small size, i.e., with less than 30,000 trainable parameters. Even when presented with data obtained from multiple operating conditions, the network is still capable to accurately predict the condition of the gearbox under inspection.
\end{abstract}
\vspace{0.2cm}

% \IEEEpeerreviewmaketitle

\section{Introduction}\label{sec:introduction}
Industrial machines are subjected to heavy stress conditions and are therefore susceptible to defects and resulting malfunctions. Even small defects can already have significant consequences as they can cause additional costs by, e.g.,  delaying industrial production through unexpected downtime, or can even put human lives in danger, e.g., resulting from bearing failure in rail vehicles. These damages can stem from normal wear and tear, from poor maintenance or completely unforeseen events. To detect such damages in an early stage whenever possible, reliable condition monitoring techniques are of utmost importance. However, the manual inspection of machines proves difficult and cost-intensive and therefore automatic non-intrusive condition monitoring techniques are preferred and highly sought after. To this end, the analysis of vibration signals has proven to be especially effective since faults especially in rotating machinery alter the typical vibration pattern of a machine. These alterations in the vibration signature can be detected by applying signal processing techniques, e.g., envelope spectrum analysis \cite{randall_rb_rolling_2011}, signal decomposition and filtering techniques \cite{peng2022}, or by manually extracting fault-related features \cite{review_feature} \cite{kreuzer2021}. However, these traditional methods have recently been eclipsed by end-to-end deep learning-based approaches \cite{neupane2020,hamadache2019,zhang2020deep,zhao2019,review_struct_health}. Although \acp{DNN} have proven to be very powerful for the task of fault detection, they come with certain drawbacks. \ac{DNN}-based approaches require a sufficiently large amount of training data for performing well. Yet, obtaining a sufficient amount of data, especially for faulty conditions is often difficult. Moreover, many architectures rely on a large number of parameters, which makes the training process and the usage with on-board devices difficult. If the number of parameters is too large and the amount of training data insufficient, these networks tend to overfit and not to generalize well for unseen test data.  To avoid these drawbacks, we propose a residual \acf{CNN} that stands out for its small number of parameters and additonally employ data augmentation techniques to further mitigate the effect of a possible over-fitting for the \ac{ICPHM} 2023 Data Challenge on Industrial Systems’ Health Monitoring using Vibration 
Signal Analysis \cite{datasheet_challenge}.

This paper is structured as follows: In Sec.~\ref{sec:dataset}, the dataset of the \ac{ICPHM} 2023 Data Challenge is described. The proposed fault detection approach is then presented in Sec.~\ref{sec:approach}. In particular, the proposed network architecture is discussed in \ref{sec:network}, whereas the utilized data augmentation techniques and the training procedure are described in Sec.~\ref{sec:data_augmentation} and Sec.~\ref{sec:training}, respectively. After this, the performance of the model is evaluated in Sec.~\ref{sec:results} before conclusions are drawn in Sec.~\ref{sec:conclusion}.

\section{Dataset}\label{sec:dataset}
\begin{figure}[htb]
    \centering
    \includegraphics[scale=0.25]{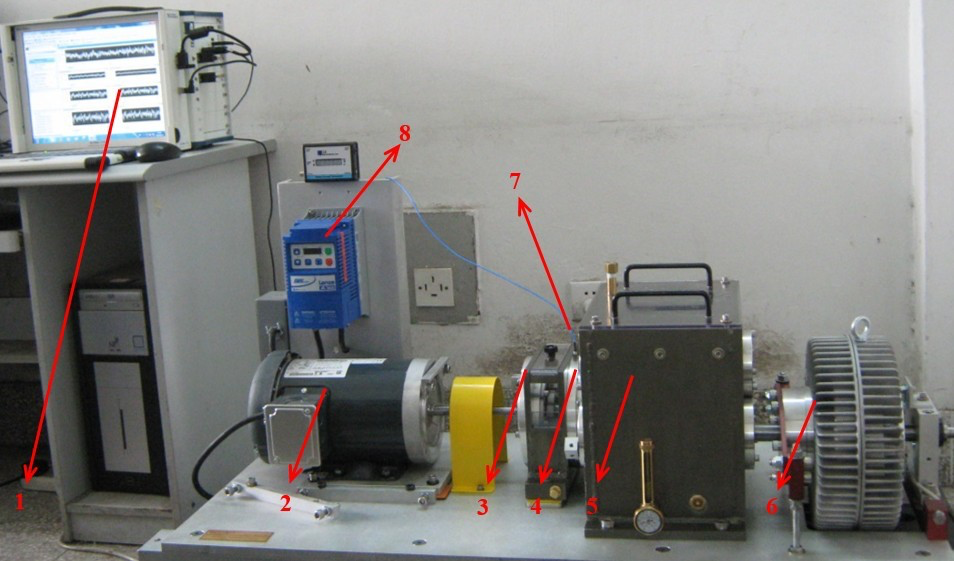}
    \caption{Test rig setup:  1. data acquisition system, 2. AC motor, 3. and 4. first stage and second stage of planetary gearbox, 5. parallel gearbox, 6. magnetic brake, 7. acceleration sensor, 8. speed transducer (taken from \cite{datasheet_challenge}).}
    \label{fig:setup}
\end{figure}

The data \cite{datasheet_challenge} was collected by the Mälardalen University (MDU)/Mälardalen Industrial Technology Center (MITC) from a test rig mounted in a lab environment. The test rig consists of an alternative current driving motor, a two-stage planetary gearbox, a two-stage parallel gearbox, and a magnetic brake, which applies torque to the output shaft of the planetary gearbox. The described setup is also shown in Fig.~\ref{fig:setup}.  The rotational speed, which can be adjusted with a speed transducer, varies in the range between 0 to 6000 revolutions per minute (rpm). In this evaluation only the planetary gearbox is considered. In particular, faults at the sun gear teeth on the second stage of the planetary gearbox are investigated. For this, an acceleration sensor, which captures vibration data in three dimensions (x, y, z) with a sampling rate $f_s = 10\,\mathrm{kHz}$, is installed on the second stage of the planetary gearbox. Further, in addition to a normal sun gear, four different common sun gear fault conditions, i.e., surface wear, chipped, cracking, and a missing tooth, are considered. The corresponding sun gears are shown in Fig.~\ref{fig:faults}. Thus, we have a classification task with $C=5$ classes with the class labels $y_i \in \{0,1,2,3,4\}$. The classes  and their corresponding labels are summarized in Tab.~\ref{tab:class_description}. For this data challenge two different operating conditions, i.e., \textit{Operating Condition 1 (OC1)} and \textit{Operating Condition 2 (OC2)}, with two different operational speeds and loads, are being examined. The specifics of these operating conditions are given in Tab.~\ref{tab:operating_conditions}.
The considered dataset is balanced as for each of the five conditions and operating conditions 10,000 frames of length 200 samples (20 ms) are available. Consequently, 50,000 frames per operating condition and 100,000 frames are available in total.

\begin{figure}[htb]
    \centering
    \input{fault_figure}
    \caption{Sun gears for each class together with a total view of the planetary gearbox (taken from \cite{datasheet_challenge}).}
    \label{fig:faults}
\end{figure}
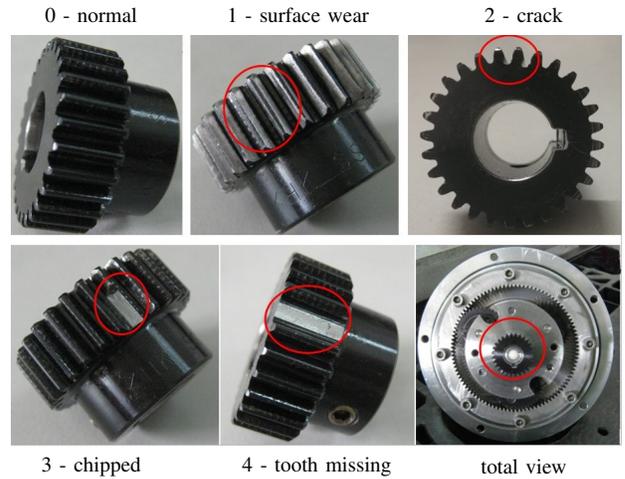
\begin{table}[htb]
    \caption{Class labels and descriptions for the five classes in the dataset.}
    \label{tab:class_description}
    \centering
    \begin{tabular}{c|c}
        \textbf{Class label} & \textbf{Description}  \\
        \hline
         0 & normal \\
         1 & surface wear \\
         2 & crack \\
         3 & chipped \\
         4 & tooth missing \\
         \hline
    \end{tabular}
    
\end{table}

\begin{table}[htb]
    \centering
     \caption{Rotational speed and load for the two operating conditions, i.e., \textit{Operating Condition 1 (OC1)} and \textit{Operating Condition 2 (OC2)}, in the training data set.}
    \label{tab:operating_conditions}
     \resizebox{\columnwidth}{!}{%
    \begin{tabular}{c| c c}
     \textbf{Operating Condition} & \textbf{Rotational Speed} [$\mathbf{\mathrm{rpm/Hz}}$]  & \textbf{Load} [$\mathbf{\mathrm{Nm}}$]  \\
      \hline
      OC1   &  1500/25 & 10 \\
    
        OC2   &  2700/45 & 25 \\
        \hline
    \end{tabular}
    }
   
\end{table}

\section{Fault Detection Approach}\label{sec:approach}
In the following the proposed fault detection approach is presented. At first, the network architecture is described in Sec.~\ref{sec:network}. Then, the data augmentation techniques that are used to introduce more diversity into the training data are presented in Sec.~\ref{sec:data_augmentation}. Sec.~\ref{sec:approach} then concludes with the description of the training procedure in Sec.~\ref{sec:training}.
\subsection{Network Architecture}\label{sec:network}
\begin{figure}
    \centering
    \input{network}
    \caption{Network architecture of the proposed model.}
    \label{fig:network}
\end{figure}
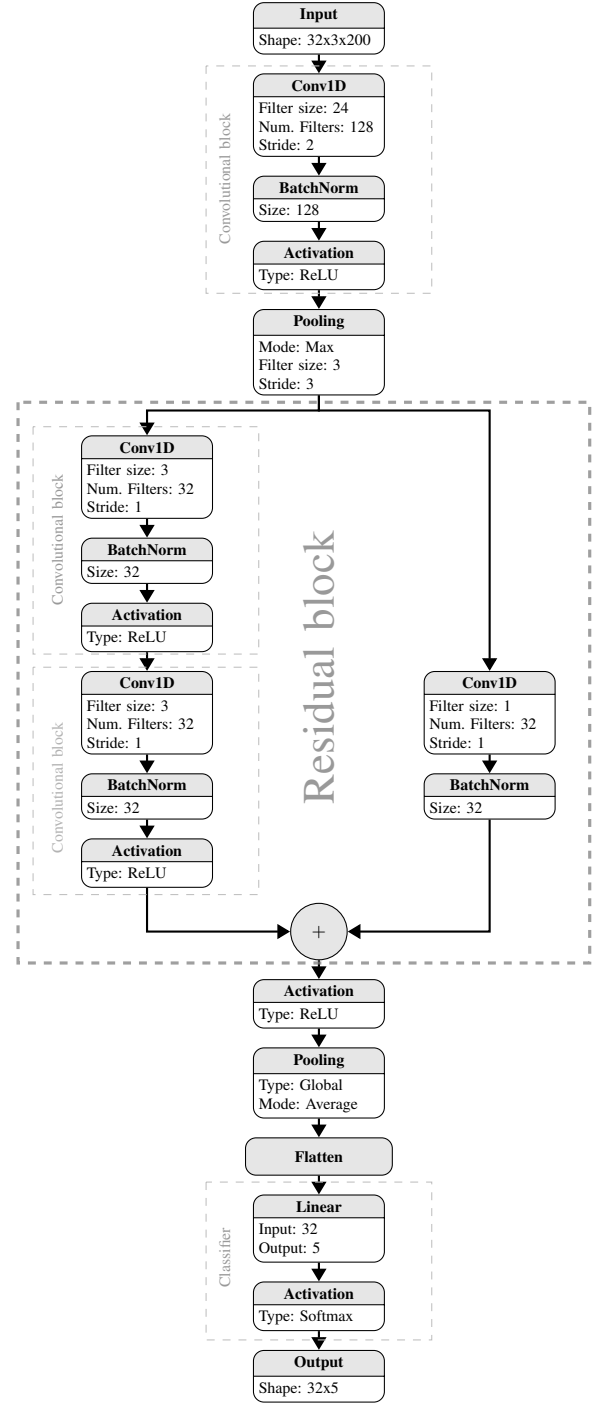

Our proposed network architecture is derived from the popular ResNet \cite{resnet}, which has proven to be very efficient for a variety of  tasks, e.g., image classification, speech recognition \cite{resnet_speech},
but also condition monitoring \cite{resnet_condition}. ResNet is a deep convolutional neural network architecture that was developed to specifically address the problem of vanishing gradients during training via backpropagation. The basic building block of a Resnet is the residual block, which consists of two or three convolutional layers with so-called shortcut or skip connections that bypass one or more layers. By employing skip connections, information is allowed to flow more easily through the network and the likelihood of vanishing gradients during training is reduced significantly \cite{resnet}. Yet, typical implementations, i.e., ResNet-34 (21.8M parameters), ResNet-50 (23.5M parameters), ResNet-152 (60.2M parameters), are very deep neural networks with 33, 49 and 151 convolutional layers, respectively. Hence, in order to train these networks from scratch  a massive amount of training data is required. Otherwise, the model is very likely to overfit. Moreover, ResNet was designed for two-dimensional image data. For these reasons, we derived a new \ac{CNN} architecture, which is well-suited for the problem at hand, i.e., the detection of faults in a planetary gearbox. The proposed architecture is depicted in Fig.~\ref{fig:network}. Note that the architecture in Fig.~\ref{fig:network} uses a batch size of 32 (cf. Sec.~\ref{sec:training}). In total, the network consists of only three one-dimensional convolutional layers and one residual block. The model uses $20\,\mathrm{ms}$ of raw time-domain vibration signals as input, which results in  $3 \times 200$-dimensional input features provided by the x-, y- and z-channel of the acceleration sensor. Furthermore, our experience with other real-world datasets has shown that using wide filters in the first layers is beneficial. This coincides with the findings in other works \cite{zhang2017,zhang2019,shouhou2022}. In particular, we use 128 filters of size 24 in the first convolutional layer, which operates along the time-axis of the input features. Each convolutional block consists of a convolutional layer that is followed by a \textit{batch normalisation} layer and a \textit{ReLU} activation function. We  also evaluated different activation functions, i.e., \textit{LeakyReLU} and \textit{Mish}\cite{mish}, but \textit{ReLU} showed the best performance. The first convolutional block is followed by a \textit{Max Pooling} layer with a filter size of 3. This first stage is followed by a residual block with two convolutional blocks with 32 filters of size 3 in the left branch. The right branch comprises a convolutional layer with filter size 1 that adjusts the dimensions of the signal for addition. After the addition of both branches and another ReLU activation, a \textit{Global Average Pooling} layer reduces the size of the feature vector to 32. A fully connected layer with 32 neurons at the input and 5 neurons at the ouput (corresponding to the number of classes) in conjunction with a \textit{Softmax} activation function acts as the classifier. The network has 29,300 trainable parameters and is trained using the cross-entropy loss function \cite{bishop2006}.
\subsection{Data Augmentation Techniques}\label{sec:data_augmentation}
To increase the robustness of the model and to thereby force the network to learn more general features, we pursue an online data augmentation approach, i.e., the samples of each batch are randomly augmented during training to increase the variety of the training samples. To this end, three different augmentation techniques are considered:  additive white gaussian noise, random amplitude scaling and circular shift. Note that these techniques are used solely on the training data and not on the validation and test data. Due to the fact that the final test set of the \ac{ICPHM} 2023 will entail data from the same operating conditions and fault classes, the data is only slightly perturbed. Therefore with a probability of $p_{\text(AWGN)}=0.25$ white gaussian noise is added to a sample in the batch such that the resulting SNR is either $25 \,\mathrm{dB}$ or $30 \, \mathrm{dB}$, as our experiments have shown that lower SNR values are not beneficial. Further with a probability of $p_{\mathrm{scale}}=0.2$, each signal is scaled with a random factor $\beta$ which is sampled from the uniform distribution:
\begin{align}
   p(\beta) = \begin{cases}
         \frac{1}{1.3-0.7} & \text{if } 0.7 \leq \beta \leq 1.3 \\
         0 & \text{otherwise}.
       \end{cases}
\end{align}
With probability $p_{\mathrm{sign}}=0.5$, the sign of $\beta$ also changes. Additionally, the samples are shifted in a circular manner with probability $p_{\mathrm{shift}}=0.2$ by an integer factor $\kappa = \lfloor \Tilde{\kappa }\rfloor $ with 
\begin{align}
   p(\Tilde{\kappa }) = \begin{cases}
         \frac{1}{200} & \text{if } 0 \leq \Tilde{\kappa } \leq 200 \\
         0 & \text{otherwise}.
       \end{cases}
\end{align}
Note that these augmentation techniques are not mutually exclusive, i.e, a single sample in the test batch could be scaled, perturbed by additive noise, and also shifted. 
\subsection{Training Procedure}\label{sec:training}
The model is evaluated using a 5-fold cross-validation approach. Hence, $20\%$ of the data are reserved for testing, while the remaining $80\%$ are used for training the model. Again, an 80/20 split is performed on the remaining four folds to generate the training and the validation set. Each split is performed in a such a manner that the number of examples for each class is identical in each partition. When the model is trained for either one of the operating conditions listed in Tab.~\ref{tab:operating_conditions}, 50,000 samples are available and thus the number of samples in the test, the training and the validation set correspond to $(N_{test},N_{train},N_{val})=(10000,32000,8000)$, which corresponds to 2000, 6400 and 1600 data samples per class label. Each batch consists of 32 samples and the weights of the network are optimized using the Adam \cite{adam} optimization algorithm with a weight decay factor of $5\times 10^{-5}$. The weight decay factor is a regularization parameter that is multiplied with the L2 norm of the weight matrix. Weight decay encourages the optimizer to keep the weights small, which can prevent over-fitting and improve generalization ability \cite{adam}. The initial learning rate is set to $\lambda = 0.001$ but is decreased by a factor of 0.8, whenever the accuracy on the validation set has not increased by at least $0.01\%$ over the last 5 epochs. To even further mitigate the risk of over-fitting, an early stopping scheme is implemented. The model is trained for a minimum number of epochs, i.e., 65 epochs, but then training is stopped when the validation accuracy has not improved over the most recent 25 epochs. The maximum number of epochs is 120. Whenever the validation accuracy increases, the network parameters are saved. The parameters of the best performing model during training are then utilized for testing the model.

\section{Results}\label{sec:results}
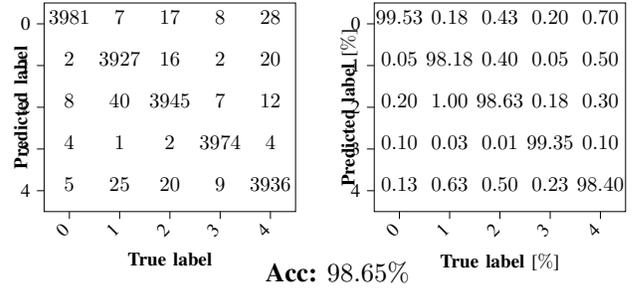
\begin{figure}[htb]
    \centering
    \input{conf_mat_5_class_model_3}
    \caption{Confusion Matrix for \textit{Model 3} Fold 1.}
    \label{fig:confmat3}
\end{figure}

The proposed model is now evaluated for three different scenarios: 
\begin{itemize}
    \item \textit{Model 1}: The model is trained and tested with data from \textit{OC1}
     \item \textit{Model 2}: The model is trained and tested with data from \textit{OC2}
     \item  \textit{Model 3}: The model is trained and tested with a balanced mixture of data   from \textit{OC1} and \textit{OC2} 
\end{itemize}
As mentioned above, all three scenarios were evaluated following a 5-fold cross-validation approach. Although \textit{Model~3} is irrelevant for the \ac{ICPHM} 2023 Data Challenge, we consider this scenario to be highly relevant since in practical scenarios, condition monitoring methods that can handle data from various operating conditions are preferable. The results for all three models are summarized in Tab.~\ref{tab:accuracy}. Tab.~\ref{tab:accuracy} shows the accuracy values for all five cross-validation folds for the three models. The mean accuracy across all folds is given in the last column for each model in Tab.~\ref{tab:accuracy}. It can be seen that the proposed model consistently predicts the condition of the planetary gearbox for all three models with a minimum accuracy of more than $98.60\%$ and an average accuracy of $98.68\%$ for the mixed-conditions model, i.e, \textit{Model~3}, and an average accuracy of more than $98.80\%$ for the single-condition models, i.e., \textit{Model~1} and \textit{Model~2}. 
The robustness of the model is evidenced by the fact that the accuracy values vary only slightly for different folds.
Furthermore, it can be noted that the difference in performance for \textit{Model~1} and \textit{Model~2} is negligibly small since the difference of the mean accuracy values is smaller than $0.09\%$. The results for \textit{Model~1} are especially remarkable since the model was trained and tested with data from \textit{OC1} for which the rotational speed is $25\,\mathrm{Hz}$. Hence, with the given frame length of $20\,\mathrm{ms}$ for the data of the challenge dataset, not even one full rotation of the sun gear is captured. This is why we strongly believe that the results could be improved even further by considering a larger frame length. Unsurprisingly, the mean accuracy for \textit{Model~3} is slightly lower as the classification tasks became more complex and challenging by considering different operating conditions with different rotational speeds and loads. Yet, the classification performance is still close to the single-condition models. It could probably be further improved by incorporating information about the operating condition into the classifier. For further illustration, the confusion matrices for Fold 1 of the models \textit{Model~1}, \textit{Model~2} and \textit{Model~3}, are shown in Fig.~\ref{fig:confmat1}, Fig.~\ref{fig:confmat2} and Fig.~\ref{fig:confmat3}, respectively. From inspecting Fig.~\ref{fig:confmat1} - Fig.~\ref{fig:confmat3}, it becomes evident that the number of false alarms, i.e., a sample is classified as a fault class although the condition is normal, is extremely low for all models as the accuracy for class label $0$ is always greater than $99.20\%$. Beyond that, it is proving difficult to derive any further generalizable insights from the confusion matrices, e.g, regularities w.r.t. misclassifications, as the predictions are very accurate in general.
\begin{table}[th!]
 \caption{Accuracy values for the three models across a 5-fold cross-validation, with mean accuracy across all folds.}
    \label{tab:accuracy}
    \centering
    \resizebox{0.95\columnwidth}{!}{%
    \begin{tabular}{|c||c|c|c|c|c||c|}
    \hline 
    \textbf{Model} & \textbf{Fold 1} & \textbf{Fold 2} & \textbf{Fold 3} & \textbf{Fold 4} & \textbf{Fold 5} & \textbf{Mean} \\
    \hline 
    \textit{Model 1} & 98.84 & 98.74 & 98.80 & 98.85 & 98.87 & 98.82 \\
    \hline
    \textit{\textit{Model 2}} & 98.90 & 98.98 & 98.80 & 98.94 & 98.94 & 98.91 \\
    \hline
    \textit{Model 3} & 98.65 & 98.73 & 98.62 & 98.63 & 98.82 & 98.68 \\
    \hline
    \end{tabular}
    }
   
\end{table}

\begin{figure}[htb]
    \centering
    \input{conf_mat_5_class_model_1}
    \caption{Confusion Matrix for \textit{Model 1} Fold 1.}
    \label{fig:confmat1}
\end{figure}
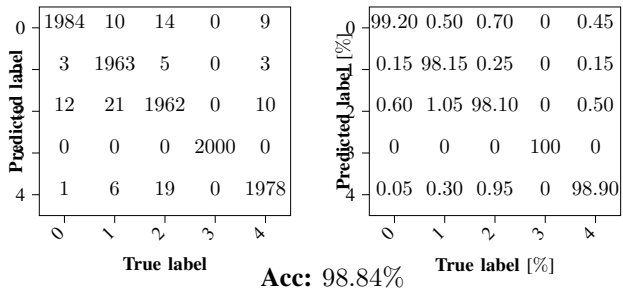
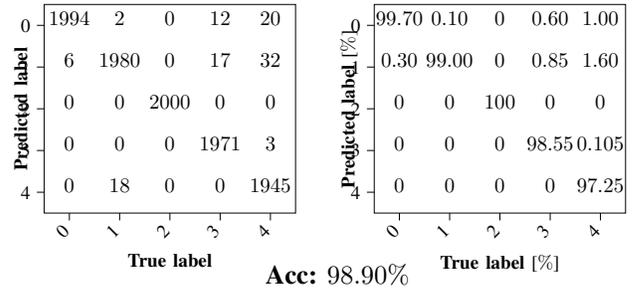
\begin{figure}[htb]
    \centering
    \input{conf_mat_5_class_model_2}
    \caption{Confusion Matrix for \textit{Model 2} Fold 1.}
    \label{fig:confmat2}
\end{figure}

\section{Conclusion}\label{sec:conclusion}
 In this article, we presented a residual-based \ac{CNN} architecture for the condition monitoring of machines   that consistently produces accuracies above $98.60\%$  for the dataset of the \ac{ICPHM} 2023 Data Challenge while using a relatively shallow architecture with less than 30K trainable parameters. It features a combination of several regularization and data augmentation techniques to reduce the risk of over-fitting and to increase the robustness of the model. The proposed model even performs well when it is subjected to data obtained from different operating conditions.

% \section*{Acknowledgment}

\bibliographystyle{IEEEtran}
\IEEEtriggeratref{13}
\bibliography{bib}
\end{document}

%% file: fault_figure.tex
\begin{tikzpicture}

\node at (0,0) {\includegraphics[scale=0.25]{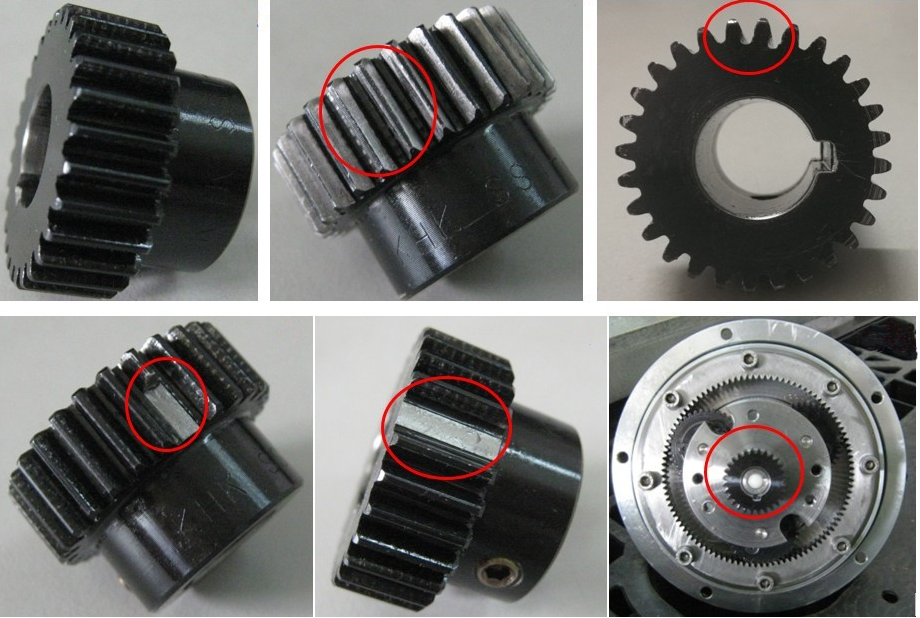}};
\node[color=black] at (-3,3) (c0) {\scriptsize 0 - normal};
\node[color=black] at (-0.25,3) (c0) {\scriptsize 1 - surface wear};
\node[color=black] at (2.75,3) (c0) {\scriptsize 2 - crack};
\node[color=black] at (-3,-3) (c0) {\scriptsize 3 - chipped};
\node[color=black] at (-0,-3) (c0) {\scriptsize 4 - tooth missing};
\node[color=black] at (2.75,-3) (c0) {\scriptsize total view};

\end{tikzpicture}

%% file: network.tex
\tikzset{
  every node/.style={
    minimum size=4ex,
    inner sep=5pt,
    font=\tiny,
  },
    normal/.style={
    minimum size=4ex,  
  },
  split/.style={
    rectangle split,
    rectangle split parts=2,
    draw,
    inner sep=2pt,
    rectangle split,
    minimum size=6ex,
    rounded corners,
  },
  splitcolored/.style={
    split,
    rectangle split part fill={gray!20, white}, % specify the fill colors for each part
  },
}
\begin{tikzpicture}%every node/.font={scale=0.25}

\node[splitcolored,text width=1.6cm] (input) at (0,0) { \hfil \textbf{Input}  \nodepart{two} Shape: 32x3x200  };

\node[splitcolored,text width=1.6cm, below= 0.25cm of input] (conv1)  { \hfil \textbf{Conv1D} \hfill \nodepart{two} Filter size: 24 \\ Num. Filters: 128 \\ Stride: 2 };

\node[splitcolored,text width=1.6cm, below= 0.25cm of conv1] (bn1)  { \hfil \textbf{BatchNorm} \nodepart{two}Size: 128};
\node[splitcolored,text width=1.6cm, below= 0.25cm of bn1] (relu1)  {\hfil \textbf{Activation} \nodepart{two}Type: ReLU};
\node[splitcolored,text width=1.6cm, below= 0.25cm of relu1] (maxpool)  {\hfil \textbf{Pooling} \nodepart{two} 
Mode: Max \\
Filter size: 3 \\ Stride: 3};

\draw[-Triangle,thick] (input) -- (conv1);
\draw[-Triangle,thick] (conv1) -- (bn1);
\draw[-Triangle,thick] (bn1) -- (relu1);
\draw[-Triangle,thick] (relu1) -- (maxpool);
\node[splitcolored,text width=1.6cm, below left = 0.75cm of maxpool] (conv2)  {  \hfil \textbf{Conv1D} \nodepart{two} Filter size: 3 \\ Num. Filters: 32 \\ Stride: 1 };

\node[splitcolored,text width=1.6cm, below= 0.25cm of conv2] (bn2)  { \hfil \textbf{BatchNorm} \nodepart{two}Size: 32};
\node[splitcolored,text width=1.6cm, below= 0.25cm of bn2] (relu2)  {\hfil \textbf{Activation} \nodepart{two}Type: ReLU};

\draw[-Triangle,thick] (maxpool.south) -- ++(0,-0.2) -| (conv2);
\draw[-Triangle,thick] (conv2) -- (bn2);
\draw[-Triangle,thick] (bn2) -- (relu2);

\node[splitcolored,text width=1.6cm, below = 0.25cm of relu2] (conv3)  { \hfil \textbf{Conv1D} \hfill \nodepart{two} Filter size: 3 \\ Num. Filters: 32 \\ Stride: 1 };

\node[splitcolored,text width=1.6cm, below= 0.25cm of conv3] (bn3)  { \hfil \textbf{BatchNorm} \nodepart{two}Size: 32};
\node[splitcolored,text width=1.6cm, below= 0.25cm of bn3] (relu3)  {\hfil \textbf{Activation} \nodepart{two}Type: ReLU};

\draw[-Triangle,thick] (relu2) -- (conv3);
\draw[-Triangle,thick] (conv3) -- (bn3);
\draw[-Triangle,thick] (bn3) -- (relu3);

\node[fill=gray!20, draw,circle, below = 6.75cm of maxpool] (add) {$+$};
\draw[-Triangle,thick] (relu3) |- (add.west);
\node[splitcolored,text width=1.6cm, right  = 2.8cm of conv3] (conv4)  {\hfil \textbf{Conv1D} \hfill \nodepart{two} Filter size: 1 \\ Num. Filters: 32 \\ Stride: 1 };
\node[splitcolored,text width=1.6cm, below= 0.25cm  of conv4] (bn4)  {\hfil\textbf{BatchNorm} \nodepart{two}Size: 32};
% \node[splitcolored,text width=1.6cm, below= 0.25cm of bn4] (relu4)  {\hfil\textbf{Activation} \nodepart{two}Type: ReLU};
\draw[-Triangle,thick] (maxpool.south) --++ (0,-0.2) -| (conv4);
\draw[-Triangle,thick] (conv4) -- (bn4);
% \draw[-Triangle,thick] (bn4) -- (relu4);
\draw[-Triangle,thick] (bn4) |- (add.east);

\node[splitcolored,text width=1.6cm, below= 0.25cm of add] (relu5)  {\hfil \textbf{Activation} \nodepart{two}Type: ReLU};
\draw[-Triangle,thick] (add) -- (relu5);
\node[splitcolored,text width=1.6cm, below= 0.25cm of relu5] (pool2)  {\hfil \textbf{Pooling} \nodepart{two}
Type: Global \\
Mode: Average};
\draw[-Triangle,thick] (relu5) -- (pool2);
\node[rectangle,rounded corners,text width=1.6cm,   draw, minimum size=1ex,align=center, fill=gray!20,
below= 0.25cm of pool2] (flatten) {\textbf{Flatten}};
\draw[-Triangle,thick] (pool2) -- (flatten);
\node[splitcolored,text width=1.6cm, below= 0.25cm of flatten] (fully)  {\hfil \textbf{Linear} \nodepart{two}Input: 32 \\ Output: 5};

\node[splitcolored,text width=1.6cm, below= 0.25cm of fully] (softmax)  {\hfil \textbf{Activation} \nodepart{two}Type: Softmax};
\draw[-Triangle,thick] (fully) -- (softmax);
\node[splitcolored,text width=1.6cm, below=0.25cm of softmax] (output) { \hfil \textbf{Output}  \nodepart{two} Shape: 32x5  };
\draw[-Triangle,thick] (flatten) -- (fully);

\draw[-Triangle,thick] (softmax) -- (output) {};
\node[normal,rotate=90,text=gray!80] (resblock) at (0,-8.5) {\LARGE{Residual block}};
\draw [draw=gray!50,dashed] (-1.5,-0.5) rectangle ++(3,-3.025);
\node[normal,rotate=90,text=gray!80] (resblock) at (-1.25,-2) { Convolutional block};
\draw [draw=gray!50,dashed] (-3.8,-5.3) rectangle ++(3,-3.025);
\node[normal,rotate=90,text=gray!80] (resblock) at (-3.475,-6.8) {Convolutional block};
\draw [draw=gray!50,dashed] (-3.8,-8.5) rectangle ++(3,-3.025);
\node[normal,rotate=90,text=gray!60] (resblock) at (-3.475,-10.1) { Convolutional block};
\draw [draw=gray,dashed,very thick, fill opacity=0.15,draw opacity =0.75] (-4,-4.975) rectangle ++(7.6,-7.46);
\draw [draw=gray!50,dashed] (-1.5,-15.35) rectangle ++(3,-2.1);
\node[normal,rotate=90,text=gray!80] (resblock) at (-1.25,-16.35) {Classifier};

\end{tikzpicture}

%% file: conf_mat_5_class_model_1.tex
\begin{tikzpicture}[scale=0.65]

\begin{axis}[
scale=0.75,
tick align=outside,
tick pos=left,
x grid style={white!69.0196078431373!black},
xlabel={\textbf{True label}},
xmin=-0.5, xmax=4.5,
xtick style={color=black},
xtick={0,1,2,3,4},
xticklabel style={rotate=45.0,yshift=0.2cm},
xticklabels={0,1,2,3,4},
y dir=reverse,
y grid style={white!69.0196078431373!black},
ylabel={\textbf{Predicted label} },
ylabel style={yshift=-0.4cm},
ymin=-0.5, ymax=4.5,
ytick style={color=black},
ytick={0,1,2,3,4},
yticklabels={0, 1,2,3,4}
]

% 1. Spalte
\draw (axis cs:0,0) node[
  scale=1.1,
  anchor=base,
  text=black,
  rotate=0.0
]{$1984$};
\draw (axis cs:0,1) node[
  scale=1.1,
  anchor=base,
  text=black,
  rotate=0.0
]{$3$};
\draw (axis cs:0,2) node[
  scale=1.1,
  anchor=base,
  text=black,
  rotate=0.0
]{$12$};
\draw (axis cs:0,3) node[
  scale=1.1,
  anchor=base,
  text=black,
  rotate=0.0
]{$0$};
\draw (axis cs:0,4) node[
  scale=1.1,
  anchor=base,
  text=black,
  rotate=0.0
]{$1$};

% 2.Spalte
\draw (axis cs:1,0) node[
  scale=1.1,
  anchor=base,
  text=black,
  rotate=0.0
]{$10$};
\draw (axis cs:1,1) node[
  scale=1.1,
  anchor=base,
  text=black,
  rotate=0.0
]{$1963$};
\draw (axis cs:1,2) node[
  scale=1.1,
  anchor=base,
  text=black,
  rotate=0.0
]{$21$};
\draw (axis cs:1,3) node[
  scale=1.1,
  anchor=base,
  text=black,
  rotate=0.0
]{$0$};
\draw (axis cs:1,4) node[
  scale=1.1,
  anchor=base,
  text=black,
  rotate=0.0
]{$6$};

% column 3
\draw (axis cs:2,0) node[
  scale=1.1,
  anchor=base,
  text=black,
  rotate=0.0
]{$14$};
\draw (axis cs:2,1) node[
  scale=1.1,
  anchor=base,
  text=black,
  rotate=0.0
]{$5$};
\draw (axis cs:2,2) node[
  scale=1.1,
  anchor=base,
  text=black,
  rotate=0.0
]{$1962$};
\draw (axis cs:2,3) node[
  scale=1.1,
  anchor=base,
  text=black,
  rotate=0.0
]{$0$};
\draw (axis cs:2,4) node[
  scale=1.1,
  anchor=base,
  text=black,
  rotate=0.0
]{$19$};
% 3rd column
\draw (axis cs:3,0) node[
  scale=1.1,
  anchor=base,
  text=black,
  rotate=0.0
]{$0$};
\draw (axis cs:3,1) node[
  scale=1.1,
  anchor=base,
  text=black,
  rotate=0.0
]{$0$};
\draw (axis cs:3,2) node[
  scale=1.1,
  anchor=base,
  text=black,
  rotate=0.0
]{$0$};
\draw (axis cs:3,3) node[
  scale=1.1,
  anchor=base,
  text=black,
  rotate=0.0
]{$2000$};
\draw (axis cs:3,4) node[
  scale=1.1,
  anchor=base,
  text=black,
  rotate=0.0
]{$0$};

\draw (axis cs:4,0) node[
  scale=1.1,
  anchor=base,
  text=black,
  rotate=0.0
]{$9$};
\draw (axis cs:4,1) node[
  scale=1.1,
  anchor=base,
  text=black,
  rotate=0.0
]{$3$};
\draw (axis cs:4,2) node[
  scale=1.1,
  anchor=base,
  text=black,
  rotate=0.0
]{$10$};
\draw (axis cs:4,3) node[
  scale=1.1,
  anchor=base,
  text=black,
  rotate=0.0
]{$0$};
\draw (axis cs:4,4) node[
  scale=1.1,
  anchor=base,
  text=black,
  rotate=0.0
]{$1978$};

\end{axis}
 \node[scale=1] at (6,-1.25) {\small \textbf{Acc: $98.84\%$}};

\begin{scope}[shift={(6.75,0)}]
\begin{axis}[
scale=0.75,
tick align=outside,
tick pos=left,
x grid style={white!69.0196078431373!black},
xlabel={\textbf{True label} [$\%$]},
xmin=-0.5, xmax=4.5,
xtick style={color=black},
xtick={0,1,2,3,4},
xticklabel style={rotate=45.0,yshift=0.2cm},
xticklabels={0,1,2,3,4},
y dir=reverse,
y grid style={white!69.0196078431373!black},
ylabel={\textbf{Predicted label} [$\%$]},
ylabel style={yshift=-0.4cm},
ymin=-0.5, ymax=4.5,
ytick style={color=black},
ytick={0,1,2,3,4},
yticklabels={0,1,2,3,4}
]

% 1. Spalte
\draw (axis cs:0,0) node[
  scale=1.1,
  anchor=base,
  text=black,
  rotate=0.0
]{$99.20$};
\draw (axis cs:0,1) node[
  scale=1.1,
  anchor=base,
  text=black,
  rotate=0.0
]{$0.15$};
\draw (axis cs:0,2) node[
  scale=1.1,
  anchor=base,
  text=black,
  rotate=0.0
]{$0.60$};
\draw (axis cs:0,3) node[
  scale=1.1,
  anchor=base,
  text=black,
  rotate=0.0
]{$0$};
\draw (axis cs:0,4) node[
  scale=1.1,
  anchor=base,
  text=black,
  rotate=0.0
]{$0.05$};

% 2.Spalte
\draw (axis cs:1,0) node[
  scale=1.1,
  anchor=base,
  text=black,
  rotate=0.0
]{$0.50$};
\draw (axis cs:1,1) node[
  scale=1.1,
  anchor=base,
  text=black,
  rotate=0.0
]{$98.15$};
\draw (axis cs:1,2) node[
  scale=1.1,
  anchor=base,
  text=black,
  rotate=0.0
]{$1.05$};
\draw (axis cs:1,3) node[
  scale=1.1,
  anchor=base,
  text=black,
  rotate=0.0
]{$0$};
\draw (axis cs:1,4) node[
  scale=1.1,
  anchor=base,
  text=black,
  rotate=0.0
]{$0.30$};
% column 3
\draw (axis cs:2,0) node[
  scale=1.1,
  anchor=base,
  text=black,
  rotate=0.0
]{$0.70$};
\draw (axis cs:2,1) node[
  scale=1.1,
  anchor=base,
  text=black,
  rotate=0.0
]{$0.25$};
\draw (axis cs:2,2) node[
  scale=1.1,
  anchor=base,
  text=black,
  rotate=0.0
]{$98.10$};
\draw (axis cs:2,3) node[
  scale=1.1,
  anchor=base,
  text=black,
  rotate=0.0
]{$0$};
\draw (axis cs:2,4) node[
  scale=1.1,
  anchor=base,
  text=black,
  rotate=0.0
]{$0.95$};

% column 3

\draw (axis cs:3,0) node[
  scale=1.1,
  anchor=base,
  text=black,
  rotate=0.0
]{$0$};
\draw (axis cs:3,1) node[
  scale=1.1,
  anchor=base,
  text=black,
  rotate=0.0
]{$0$};
\draw (axis cs:3,2) node[
  scale=1.1,
  anchor=base,
  text=black,
  rotate=0.0
]{$0$};
\draw (axis cs:3,3) node[
  scale=1.1,
  anchor=base,
  text=black,
  rotate=0.0
]{$100$};
\draw (axis cs:3,4) node[
  scale=1.1,
  anchor=base,
  text=black,
  rotate=0.0
]{$0$};

\draw (axis cs:4,0) node[
  scale=1.1,
  anchor=base,
  text=black,
  rotate=0.0
]{$0.45$};
\draw (axis cs:4,1) node[
  scale=1.1,
  anchor=base,
  text=black,
  rotate=0.0
]{$0.15$};
\draw (axis cs:4,2) node[
  scale=1.1,
  anchor=base,
  text=black,
  rotate=0.0
]{$0.50$};
\draw (axis cs:4,3) node[
  scale=1.1,
  anchor=base,
  text=black,
  rotate=0.0
]{$0$};
\draw (axis cs:4,4) node[
  scale=1.1,
  anchor=base,
  text=black,
  rotate=0.0
]{$98.90$};
\end{axis}
\end{scope}
\end{tikzpicture}

%% file: conf_mat_5_class_model_2.tex
\begin{tikzpicture}[scale=0.65]

\begin{axis}[
scale=0.75,
tick align=outside,
tick pos=left,
x grid style={white!69.0196078431373!black},
xlabel={\textbf{True label}},
xmin=-0.5, xmax=4.5,
xtick style={color=black},
xtick={0,1,2,3,4},
xticklabel style={rotate=45.0,yshift=0.2cm},
xticklabels={0,1,2,3,4},
y dir=reverse,
y grid style={white!69.0196078431373!black},
ylabel={\textbf{Predicted label} },
ylabel style={yshift=-0.4cm},
ymin=-0.5, ymax=4.5,
ytick style={color=black},
ytick={0,1,2,3,4},
yticklabels={0, 1,2,3,4}
]

% 1. Spalte
\draw (axis cs:0,0) node[
  scale=1.1,
  anchor=base,
  text=black,
  rotate=0.0
]{$1994$};
\draw (axis cs:0,1) node[
  scale=1.1,
  anchor=base,
  text=black,
  rotate=0.0
]{$6$};
\draw (axis cs:0,2) node[
  scale=1.1,
  anchor=base,
  text=black,
  rotate=0.0
]{$0$};
\draw (axis cs:0,3) node[
  scale=1.1,
  anchor=base,
  text=black,
  rotate=0.0
]{$0$};
\draw (axis cs:0,4) node[
  scale=1.1,
  anchor=base,
  text=black,
  rotate=0.0
]{$0$};

% 2.Spalte
\draw (axis cs:1,0) node[
  scale=1.1,
  anchor=base,
  text=black,
  rotate=0.0
]{$2$};
\draw (axis cs:1,1) node[
  scale=1.1,
  anchor=base,
  text=black,
  rotate=0.0
]{$1980$};
\draw (axis cs:1,2) node[
  scale=1.1,
  anchor=base,
  text=black,
  rotate=0.0
]{$0$};
\draw (axis cs:1,3) node[
  scale=1.1,
  anchor=base,
  text=black,
  rotate=0.0
]{$0$};
\draw (axis cs:1,4) node[
  scale=1.1,
  anchor=base,
  text=black,
  rotate=0.0
]{$18$};

% column 3
\draw (axis cs:2,0) node[
  scale=1.1,
  anchor=base,
  text=black,
  rotate=0.0
]{$0$};
\draw (axis cs:2,1) node[
  scale=1.1,
  anchor=base,
  text=black,
  rotate=0.0
]{$0$};
\draw (axis cs:2,2) node[
  scale=1.1,
  anchor=base,
  text=black,
  rotate=0.0
]{$2000$};
\draw (axis cs:2,3) node[
  scale=1.1,
  anchor=base,
  text=black,
  rotate=0.0
]{$0$};
\draw (axis cs:2,4) node[
  scale=1.1,
  anchor=base,
  text=black,
  rotate=0.0
]{$0$};
% 3rd column
\draw (axis cs:3,0) node[
  scale=1.1,
  anchor=base,
  text=black,
  rotate=0.0
]{$12$};
\draw (axis cs:3,1) node[
  scale=1.1,
  anchor=base,
  text=black,
  rotate=0.0
]{$17$};
\draw (axis cs:3,2) node[
  scale=1.1,
  anchor=base,
  text=black,
  rotate=0.0
]{$0$};
\draw (axis cs:3,3) node[
  scale=1.1,
  anchor=base,
  text=black,
  rotate=0.0
]{$1971$};
\draw (axis cs:3,4) node[
  scale=1.1,
  anchor=base,
  text=black,
  rotate=0.0
]{$0$};

\draw (axis cs:4,0) node[
  scale=1.1,
  anchor=base,
  text=black,
  rotate=0.0
]{$20$};
\draw (axis cs:4,1) node[
  scale=1.1,
  anchor=base,
  text=black,
  rotate=0.0
]{$32$};
\draw (axis cs:4,2) node[
  scale=1.1,
  anchor=base,
  text=black,
  rotate=0.0
]{$0$};
\draw (axis cs:4,3) node[
  scale=1.1,
  anchor=base,
  text=black,
  rotate=0.0
]{$3$};
\draw (axis cs:4,4) node[
  scale=1.1,
  anchor=base,
  text=black,
  rotate=0.0
]{$1945$};

\end{axis}
 \node[scale=1] at (6,-1.25) {\small \textbf{Acc: $98.90\%$}};

\begin{scope}[shift={(6.75,0)}]
\begin{axis}[
scale=0.75,
tick align=outside,
tick pos=left,
x grid style={white!69.0196078431373!black},
xlabel={\textbf{True label} [$\%$]},
xmin=-0.5, xmax=4.5,
xtick style={color=black},
xtick={0,1,2,3,4},
xticklabel style={rotate=45.0,yshift=0.2cm},
xticklabels={0,1,2,3,4},
y dir=reverse,
y grid style={white!69.0196078431373!black},
ylabel={\textbf{Predicted label} [$\%$]},
ylabel style={yshift=-0.4cm},
ymin=-0.5, ymax=4.5,
ytick style={color=black},
ytick={0,1,2,3,4},
yticklabels={0,1,2,3,4}
]

% 1. Spalte
\draw (axis cs:0,0) node[
  scale=1.1,
  anchor=base,
  text=black,
  rotate=0.0
]{$99.70$};
\draw (axis cs:0,1) node[
  scale=1.1,
  anchor=base,
  text=black,
  rotate=0.0
]{$0.30$};
\draw (axis cs:0,2) node[
  scale=1.1,
  anchor=base,
  text=black,
  rotate=0.0
]{$0$};
\draw (axis cs:0,3) node[
  scale=1.1,
  anchor=base,
  text=black,
  rotate=0.0
]{$0$};
\draw (axis cs:0,4) node[
  scale=1.1,
  anchor=base,
  text=black,
  rotate=0.0
]{$0$};

% 2.Spalte
\draw (axis cs:1,0) node[
  scale=1.1,
  anchor=base,
  text=black,
  rotate=0.0
]{$0.10$};
\draw (axis cs:1,1) node[
  scale=1.1,
  anchor=base,
  text=black,
  rotate=0.0
]{$99.00$};
\draw (axis cs:1,2) node[
  scale=1.1,
  anchor=base,
  text=black,
  rotate=0.0
]{$0$};
\draw (axis cs:1,3) node[
  scale=1.1,
  anchor=base,
  text=black,
  rotate=0.0
]{$0$};
\draw (axis cs:1,4) node[
  scale=1.1,
  anchor=base,
  text=black,
  rotate=0.0
]{$0$};
% column 3
\draw (axis cs:2,0) node[
  scale=1.1,
  anchor=base,
  text=black,
  rotate=0.0
]{$0$};
\draw (axis cs:2,1) node[
  scale=1.1,
  anchor=base,
  text=black,
  rotate=0.0
]{$0$};
\draw (axis cs:2,2) node[
  scale=1.1,
  anchor=base,
  text=black,
  rotate=0.0
]{$100$};
\draw (axis cs:2,3) node[
  scale=1.1,
  anchor=base,
  text=black,
  rotate=0.0
]{$0$};
\draw (axis cs:2,4) node[
  scale=1.1,
  anchor=base,
  text=black,
  rotate=0.0
]{$0$};

% column 3

\draw (axis cs:3,0) node[
  scale=1.1,
  anchor=base,
  text=black,
  rotate=0.0
]{$0.60$};
\draw (axis cs:3,1) node[
  scale=1.1,
  anchor=base,
  text=black,
  rotate=0.0
]{$0.85$};
\draw (axis cs:3,2) node[
  scale=1.1,
  anchor=base,
  text=black,
  rotate=0.0
]{$0$};
\draw (axis cs:3,3) node[
  scale=1.1,
  anchor=base,
  text=black,
  rotate=0.0
]{$98.55$};
\draw (axis cs:3,4) node[
  scale=1.1,
  anchor=base,
  text=black,
  rotate=0.0
]{$0$};

\draw (axis cs:4,0) node[
  scale=1.1,
  anchor=base,
  text=black,
  rotate=0.0
]{$1.00$};
\draw (axis cs:4,1) node[
  scale=1.1,
  anchor=base,
  text=black,
  rotate=0.0
]{$1.60$};
\draw (axis cs:4,2) node[
  scale=1.1,
  anchor=base,
  text=black,
  rotate=0.0
]{$0$};
\draw (axis cs:4,3) node[
  scale=1.1,
  anchor=base,
  text=black,
  rotate=0.0
]{$0.105$};
\draw (axis cs:4,4) node[
  scale=1.1,
  anchor=base,
  text=black,
  rotate=0.0
]{$97.25$};
\end{axis}
\end{scope}
\end{tikzpicture}

%% file: conf_mat_5_class_model_3.tex
\begin{tikzpicture}[scale=0.65]

\begin{axis}[
scale=0.75,
tick align=outside,
tick pos=left,
x grid style={white!69.0196078431373!black},
xlabel={\textbf{True label}},
xmin=-0.5, xmax=4.5,
xtick style={color=black},
xtick={0,1,2,3,4},
xticklabel style={rotate=45.0,yshift=0.2cm},
xticklabels={0,1,2,3,4},
y dir=reverse,
y grid style={white!69.0196078431373!black},
ylabel={\textbf{Predicted label} },
ylabel style={yshift=-0.4cm},
ymin=-0.5, ymax=4.5,
ytick style={color=black},
ytick={0,1,2,3,4},
yticklabels={0, 1,2,3,4}
]

% 1. Spalte
\draw (axis cs:0,0) node[
  scale=1.1,
  anchor=base,
  text=black,
  rotate=0.0
]{$3981$};
\draw (axis cs:0,1) node[
  scale=1.1,
  anchor=base,
  text=black,
  rotate=0.0
]{$2$};
\draw (axis cs:0,2) node[
  scale=1.1,
  anchor=base,
  text=black,
  rotate=0.0
]{$8$};
\draw (axis cs:0,3) node[
  scale=1.1,
  anchor=base,
  text=black,
  rotate=0.0
]{$4$};
\draw (axis cs:0,4) node[
  scale=1.1,
  anchor=base,
  text=black,
  rotate=0.0
]{$5$};

% 2.Spalte
\draw (axis cs:1,0) node[
  scale=1.1,
  anchor=base,
  text=black,
  rotate=0.0
]{$7$};
\draw (axis cs:1,1) node[
  scale=1.1,
  anchor=base,
  text=black,
  rotate=0.0
]{$3927$};
\draw (axis cs:1,2) node[
  scale=1.1,
  anchor=base,
  text=black,
  rotate=0.0
]{$40$};
\draw (axis cs:1,3) node[
  scale=1.1,
  anchor=base,
  text=black,
  rotate=0.0
]{$1$};
\draw (axis cs:1,4) node[
  scale=1.1,
  anchor=base,
  text=black,
  rotate=0.0
]{$25$};

% column 3
\draw (axis cs:2,0) node[
  scale=1.1,
  anchor=base,
  text=black,
  rotate=0.0
]{$17$};
\draw (axis cs:2,1) node[
  scale=1.1,
  anchor=base,
  text=black,
  rotate=0.0
]{$16$};
\draw (axis cs:2,2) node[
  scale=1.1,
  anchor=base,
  text=black,
  rotate=0.0
]{$3945$};
\draw (axis cs:2,3) node[
  scale=1.1,
  anchor=base,
  text=black,
  rotate=0.0
]{$2$};
\draw (axis cs:2,4) node[
  scale=1.1,
  anchor=base,
  text=black,
  rotate=0.0
]{$20$};
% 3rd column
\draw (axis cs:3,0) node[
  scale=1.1,
  anchor=base,
  text=black,
  rotate=0.0
]{$8$};
\draw (axis cs:3,1) node[
  scale=1.1,
  anchor=base,
  text=black,
  rotate=0.0
]{$2$};
\draw (axis cs:3,2) node[
  scale=1.1,
  anchor=base,
  text=black,
  rotate=0.0
]{$7$};
\draw (axis cs:3,3) node[
  scale=1.1,
  anchor=base,
  text=black,
  rotate=0.0
]{$3974$};
\draw (axis cs:3,4) node[
  scale=1.1,
  anchor=base,
  text=black,
  rotate=0.0
]{$9$};

\draw (axis cs:4,0) node[
  scale=1.1,
  anchor=base,
  text=black,
  rotate=0.0
]{$28$};
\draw (axis cs:4,1) node[
  scale=1.1,
  anchor=base,
  text=black,
  rotate=0.0
]{$20$};
\draw (axis cs:4,2) node[
  scale=1.1,
  anchor=base,
  text=black,
  rotate=0.0
]{$12$};
\draw (axis cs:4,3) node[
  scale=1.1,
  anchor=base,
  text=black,
  rotate=0.0
]{$4$};
\draw (axis cs:4,4) node[
  scale=1.1,
  anchor=base,
  text=black,
  rotate=0.0
]{$3936$};

\end{axis}
 \node[scale=1] at (6,-1.25) {\small \textbf{Acc: $98.65\%$}};

\begin{scope}[shift={(6.75,0)}]
\begin{axis}[
scale=0.75,
tick align=outside,
tick pos=left,
x grid style={white!69.0196078431373!black},
xlabel={\textbf{True label} [$\%$]},
xmin=-0.5, xmax=4.5,
xtick style={color=black},
xtick={0,1,2,3,4},
xticklabel style={rotate=45.0,yshift=0.2cm},
xticklabels={0,1,2,3,4},
y dir=reverse,
y grid style={white!69.0196078431373!black},
ylabel={\textbf{Predicted label} [$\%$]},
ylabel style={yshift=-0.4cm},
ymin=-0.5, ymax=4.5,
ytick style={color=black},
ytick={0,1,2,3,4},
yticklabels={0,1,2,3,4}
]

% 1. Spalte
\draw (axis cs:0,0) node[
  scale=1.1,
  anchor=base,
  text=black,
  rotate=0.0
]{$99.53$};
\draw (axis cs:0,1) node[
  scale=1.1,
  anchor=base,
  text=black,
  rotate=0.0
]{$0.05$};
\draw (axis cs:0,2) node[
  scale=1.1,
  anchor=base,
  text=black,
  rotate=0.0
]{$0.20$};
\draw (axis cs:0,3) node[
  scale=1.1,
  anchor=base,
  text=black,
  rotate=0.0
]{$0.10$};
\draw (axis cs:0,4) node[
  scale=1.1,
  anchor=base,
  text=black,
  rotate=0.0
]{$0.13$};

% 2.Spalte
\draw (axis cs:1,0) node[
  scale=1.1,
  anchor=base,
  text=black,
  rotate=0.0
]{$0.18$};
\draw (axis cs:1,1) node[
  scale=1.1,
  anchor=base,
  text=black,
  rotate=0.0
]{$98.18$};
\draw (axis cs:1,2) node[
  scale=1.1,
  anchor=base,
  text=black,
  rotate=0.0
]{$1.00$};
\draw (axis cs:1,3) node[
  scale=1.1,
  anchor=base,
  text=black,
  rotate=0.0
]{$0.03$};
\draw (axis cs:1,4) node[
  scale=1.1,
  anchor=base,
  text=black,
  rotate=0.0
]{$0.63$};
% column 3
\draw (axis cs:2,0) node[
  scale=1.1,
  anchor=base,
  text=black,
  rotate=0.0
]{$0.43$};
\draw (axis cs:2,1) node[
  scale=1.1,
  anchor=base,
  text=black,
  rotate=0.0
]{$0.40$};
\draw (axis cs:2,2) node[
  scale=1.1,
  anchor=base,
  text=black,
  rotate=0.0
]{$98.63$};
\draw (axis cs:2,3) node[
  scale=1.1,
  anchor=base,
  text=black,
  rotate=0.0
]{$0.01$};
\draw (axis cs:2,4) node[
  scale=1.1,
  anchor=base,
  text=black,
  rotate=0.0
]{$0.50$};

% column 3

\draw (axis cs:3,0) node[
  scale=1.1,
  anchor=base,
  text=black,
  rotate=0.0
]{$0.20$};
\draw (axis cs:3,1) node[
  scale=1.1,
  anchor=base,
  text=black,
  rotate=0.0
]{$0.05$};
\draw (axis cs:3,2) node[
  scale=1.1,
  anchor=base,
  text=black,
  rotate=0.0
]{$0.18$};
\draw (axis cs:3,3) node[
  scale=1.1,
  anchor=base,
  text=black,
  rotate=0.0
]{$99.35$};
\draw (axis cs:3,4) node[
  scale=1.1,
  anchor=base,
  text=black,
  rotate=0.0
]{$0.23$};

\draw (axis cs:4,0) node[
  scale=1.1,
  anchor=base,
  text=black,
  rotate=0.0
]{$0.70$};
\draw (axis cs:4,1) node[
  scale=1.1,
  anchor=base,
  text=black,
  rotate=0.0
]{$0.50$};
\draw (axis cs:4,2) node[
  scale=1.1,
  anchor=base,
  text=black,
  rotate=0.0
]{$0.30$};
\draw (axis cs:4,3) node[
  scale=1.1,
  anchor=base,
  text=black,
  rotate=0.0
]{$0.10$};
\draw (axis cs:4,4) node[
  scale=1.1,
  anchor=base,
  text=black,
  rotate=0.0
]{$98.40$};
\end{axis}
\end{scope}
\end{tikzpicture}